# Enabling Blockchain Scalability and Interoperability with Mobile Computing through LayerOne.X

**Kevin Coutinho, Alphabright Digital**

**Ponnie Clark, University of Western Australia**

**Ferdinand Azis, Alphabright Digital**

**Norman Lip, Alphabright Digital**

**Josh Hunt, Alphabright Digital**

Interoperability and scalability are currently the bottlenecks preventing mass adoption of blockchain technology. Development of an interoperable and scalable network that promotes a truly decentralised, permissionless and secure blockchain as well as one that enables micro validation is the main goal of this project. Layer-One.X, a truly decentralised ledger which utilises para-sharding, Directed Acyclic Graphs, Proof of Participation consensus mechanism, mobile computing, flash contracts and nucleus scripting is introduced in this paper. The conceptual framework including tokenomics is also explained along with a number of use cases. The framework facilitates the growing need of transaction per second enabling micro based payments and value transfer through tokenisation.

## 1. Introduction

Blockchain technology has maintained significant interest from both industry and academia. The technology is progressing at a rapid pace. However, adoption of blockchain technology has been limited due to challenges over scalability, interoperability, and privacy while keeping the blockchain trilemma uncompromised for decentralisation and security. In this project, we propose a Layer-One.X solution framework, designed to address the scalability and interoperability challenges. It is designed to support scalability and provide a trustless mechanism for cross chain (homogeneous blockchains) and cross blockchain (heterogeneous blockchains) interoperability. Development of an interoperable and scalable network that promotes a truly decentralised, permissionless, and secure blockchain is the main goal of this project, enabling micro validation and tokenisation.

Decentralised tamper resistance is one of the key technologies of blockchain and has a broad application prospect. However, blockchain scalability bottlenecks affect mainstream adoption due to the blockchain trilemma of decentralisation, security, and scalability. A number of solutions have been proposed to overcome scalability issues, which can be classified into five main streams; sharding, Directed Acyclic Graph (DAG), off-chain, cross chain, and edge computing technologies.

These existing solutions offer alternate approaches but do not directly solve the problem. For example, a method aims to reduce the latency from the users' perspective but does not improve the throughput of blockchains [1]. As a result, scalability is still challenging in blockchain. In this project, we propose the Layer-One.Scalability (Layer-One.S) solution to tackle the scalability issues. One approach of blockchain scalability solutions is to provide a way to offload transactions to other blockchains but these blockchains need to be interoperable. The problem is that current blockchains are not intrinsically designed to be interoperable. Each blockchain is focused on resolving particular challenges within industries in a siloed manner. The interoperability needs are overlooked, leading to data and value silos [2]–[4], thereby becoming obsolete and vulnerable, and in a worst-case scenario, resulting in shutdown stage. There are attempts to develop approaches and protocols tackling blockchain interoperability limitations however assumptions and compromises are required for security and decentralisation. Several surveys on blockchain interoperability conclude that fully decentralised blockchain interoperability is feasible only with the presence of a trusted third party [5]. Interoperability functions such as asset transfer are not possible without a third party's mediation to assure the correctness of the underlying protocol [6]. In this paper, we discuss the possibility of connecting homogeneous and heterogeneous blockchains through a layer one solution called Layer-One.Interoperability (Layer-One.I).

The Layer-One.X solution enables (i) interoperating between any public and private blockchains for data and asset value exchange, (ii) scaling for fast transaction processing and to increase data capacity to blockchains, and (iii) an economic system that is beneficial from participating in the ecosystems. Layer-One.X is a state-of-the-art blockchain that introduces a new paradigm of scalability and interoperability.

The remainder of this paper is structured as follows. In the next section, section 2, literature review on the existing scalability solutions and the challenges presented, followed by a discussion on the Layer-One.S solution. In section 3, the existing interoperability solutions and challenges are discussed, followed by a discussion on the Layer-One.I solution. In section 4, token economies are reviewed and the tokenised economic for the Layer-One.X system is defined. In Section 5, use cases of the Layer-One.X system are discussed. The paper is concluded in section 5.

## 2. Scalability

Blockchain still faces scalability issues. In this section, the five mainstream solutions aiming to improve the performance of blockchain scalability are reviewed. These solutions are sharding, Directed Acyclic Graph (DAG), off chain, cross chain, and edge computing technologies. Research direction of and into the Layer-One.S solution is discussed subsequently.

### 2.1 Review of scalability issues

<u>Sharding</u>
Sharding is one of the most practical solutions in achieving a scalable system while also increasing network size. For sharding, the network is partitioned into different groups, i.e. multiple shards, thereby the overhead of duplicating communication, storage, and computation in each participating node can be reduced [7]. In traditional blockchain, these overheads are incurred by all full nodes. With the sharding method, the processing, storage, and computing can be conducted in parallel. The capacity and throughput become linearly proportionate to the number of participating nodes and the number of shards respectively. Hence decentralisation and security are preserved while achieving scalability. There are a number of projects working on different sharding mechanisms as set out in Table 1.

Directed Acyclic Graph (DAG)
DAG as a distributed ledger technology has been used to resolve speed and scalability issues with low transaction costs. As opposed to a traditional blockchain, DAG technology is an alternative approach, differing in how transactions are added into a network. Individual DAG transactions are linked to one another forming a graph rather than connected together in a chain of blocks [8]. Since a DAG structure can position many blocks at once, the transaction speed (per second) is higher than chaining blocks. Many see DAG technology as the next evolution of blockchain due to its novel structures that may potentially support scalability. However, many see DAG as truly not a blockchain, given that the terms "Distributed Ledger Technology" (DLT) and "Blockchain Technology" cannot be used synonymously. DLT is a broader term and DAG is another type of DLT. There are a number of projects developed with different DAG protocols and designs as set out in Table 1.

Off Chain
Off chain solution is another approach trying to improve the performance of blockchain systems. In blockchain-based solutions, data has been managed either on-chain or off-chain as storage mechanisms. Off-chain means that data is on the blockchain but not on publicly accessible service. Off chain faces problems with synchronising finality on an instant transaction basis. Off-chain can be implemented with the traditional centralised distributed system and the performance is scalable. Using this strategy, off-chain transactions only broadcast a summary of a batch of off-chain transactions to the main blockchain [9]. Truly fine-grained details of bilateral transactions are not recorded on the main blockchain [10].

Cross Chain
The challenge of scalability and interoperability are indeed interrelated in many ways. One solution for both challenges is to develop technology that takes pressure off the main blockchain by allowing it to operate in another blockchain and facilitates the interactions between different blockchains. The solution known as cross chain technology enhances scalability by leveraging the combined respective throughputs of blockchains. As such, cross-chain transactions and smart contracts can be managed on different blockchain networks. Speed can be leveraged from multiple chains enhancing scalability. A number of projects in off-chain and cross chain technologies are set out in Table 1.

Mobile Computing
The integration of blockchain with mobile computing has been the subject of recent research and can be categorised into three groups i.e. (i) mobile computing as part of blockchain e.g. those in [11]–[13], (ii) blockchain to enhance mobile edge services e.g. those in [14]–[19], and (iii) blockchain to manage mobile edge resources e.g. those in [16], [18], [20]–[22]. For this project, the first group is the focus. One of the critical points is resource orchestration [23]. Mobile computing infrastructure can support real time resource management in which computation can be offloaded through the mobile network. Identity management and resource allocation processes can be implemented. Smart contracts can manage network bandwidth and resource allocation in a distributed and collaborative computational offloading framework [19]. Optimisation algorithms to orchestrate resource allocation processes can be developed for computational offloading in a traditional deterministic model, reinforcement learning model [24], adaptive genetic model [25].

Table 1. Existing scalability solutions, against Layer-One.S solution

| Scalability project | Mechanism | Consensus Protocol | Description | Limitations |
|---|---|---|---|---|
| **Elastico** [26] | Sharding | Practical Byzantine Fault Tolerance (PBFT) | Allocating potential validators based on the least-significant bits of the result of PoW puzzles [7]. | Weak robustness as low byzantine resilience [10]. Latency scales linearly [7]. |
| **Zilliqa** [27] | Sharding | Proof of work (PoW) | Comprising shards and a directory service, where the role of each node is assigned by PoW [10]. All the nodes participate is validated by the committee, to be assigned to shards [27]. | When the number of nodes goes up, its efficiency goes down because every node has to communicate with every other node of the same shard [10]. |
| **Omniledger** [28] | Sharding | ByzCoinX | Scalability is improved with a lower failure probability for ByzCoinX consensus protocol by sacrificing the transaction latency in large scale network [28]. | Time consuming bootstrap process [10]. Prone to Denial-of-Service (DoS) attacks [7]. |
| **Monoxide** [29] | Sharding | Proof of work based Chu-ko-nu mining | Using Chu-ko-nu mining protocol to achieve a large-scale network with a huge number of shards and a flexible shard size [29]. | Rely on Chu-ko-nu mining posing the issue of power centralisation and additional overhead to Monoxide [7]. |
| **RapidChain** [30] | Sharding | 50% Byzantine Fault Tolerance (BFT) with pipelining | Runs an autonomous pre-scheduled scheme within a small sized shard to agree on a timeout based on which the consensus speed can be adjusted to prevent the asynchronisation [30]. | Suitable for small sized shards, not scalable with communication overhead. The pre-scheduled scheme defining the timeout is not credibly proved for synchronisation [7]. |

| Scalability project | Mechanism | Consensus Protocol | Description | Limitations |
|---|---|---|---|---|
| **Chainspace** [31] | Sharding | Implementation of PBFT: MOD-SMART | Introduces an audit scheme to prevent attacks from dishonest shards. It allocates nodes in different shards; the elected leader of the shard takes responsibility for the atomicity of cross-shard transactions [31]. | Sacrificing smart-contract-oriented sharding mechanism with the cost of overhead and low throughput [7]. |
| **Ethereum 2.0** [32] | Sharding | BFT based PoS | Selecting the proposer of each shard followed by selecting attesters for each shard and the validators responsible for checkpointing from the global pool are selected [33]. | Still in experimental stage. Casper incurs large overhead. Randomness generator incurs high computational overhead depending on the incentive scheme. Prone to censorship attack and grinding attacks if the attack cost is acceptable [7]. |
| **EOS** [34] | Sharding | Distributed proof of stake | The PoS decision is taken by a restricted group of 21 producers, while all the users who have stakes on the chain participate in the election of the producers [34]. | Rely on voting hence it is critical for network security and not robust [35]. |
| **Blockclique** [36] | DAG | Bitrate | Divide nodes into threads which work in parallel, producing blocks that are organised in a DAG [36]. | Trade-off between security and efficiency. |
| **Byteball** [37], **DagCoin** [38] | DAG | Witness | Construct DAG with global transactions, not blocks. It identifies main chain relying on witness, the trustable honest nodes to achieve consensus [10]. | Trade-off between scalability and centralization |
| **Nano** [39] | DAG | Delegated proof of stake | Uses a block-lattice structure where each node has their own local blockchain [39]. | Trade-off between decentralisation and sybil resistance [10]. |

| Scalability project | Mechanism | Consensus Protocol | Description | Limitations |
|---|---|---|---|---|
| **IOTA** [13] | DAG | Tangle | The group issuing transactions is the same group confirming transactions. It offers a high level of scalability by creating blocks that do not contain whole transactions [13]. | Trade-off between latency and security. |
| **Lightning network** [40] | Offchain | Proof of work for on-chain transaction | Two-way payment channel in which users can conduct seamless transactions off chain and do not require a global consensus [40]. | Rely on a large number of multi-signature wallets hence not suitable for large payment. Full scale for privacy and security is lacking. Recurring open-closure payment channels involving on-chain transactions can be a problem causing high transaction costs [10]. |
| **Raiden network** [41] | Offchain | Proof of stake | Support off-chain payments and broadcast a summary of a batch of off-chain payments to the Blockchain [41]. | Privacy and security problem requires both parties to remain online or lock tokens for committing transactions [10]. |
| **Sprites** [42] | Offchain | Proof of stake | A user provides cross-channel payments to others through the amount of money saved in the channel with fee incentivised for the user to pay cross channel [42]. | The funds used for payment are frozen during the process of cross-channel payment [10]. |
| **Plasma** [43] | Offchain | Fraud proofs | It connects the sub-chains to the main chain through fraud proofs [43]. Also known as blockchains in blockchains forming chain trees. | Minimize the interaction with the Blockchain to reduce the latency from the users' perspective but does not improve the throughput of Blockchains [10]. |
| **Multi-centre witness** [44] | Cross chain | Witness | Uses witness to guarantee the locking and unlocking of assets in different chain [44]. | Trade-off between scalability and centralization |
| **Side chain / Relay** [45] | Cross chain | | Anchoring various chains on the main chain. | There could be a dispute occurred at closure phrase. |

| Scalability project | Mechanism | Consensus Protocol | Description | Limitations |
|---|---|---|---|---|
| **Hash locking** [44] | Cross chain | (not identify) | Using hash preimage as secret and conditional payment; the atomicity of different transactions can be guaranteed [44]. | Leading to funds being locked. |
| **Distributed private key control** [46] | Cross chain | (not identify) | Lock-in and lock out protocol is used to implement distributed control management and asset mapping of all digital assets controlled by private keys [46]. | Both sender and receiver know the private key used for asset transfer. |
| **EdgeChain** [14] | Mobile computing | PBFT | Relies on smart contracts for managing resources allocation in a distributed and collaborative computational offloading framework for blockchain [14]. | Not supporting resource provisioning for multiple heterogeneous applications [14]. Highly experimental products. |
| **Edge AI** [12] | Mobile computing | PoW | Ethereum Blockchain based architecture with edge artificial intelligence to analyse data at the edge of the network and keep track of the parties that access the results of the analysis, which are stored in distributed databases [12]. | Trade-off between latency and security. |
| **LayerOne.S** | Para-sharding, DAG, and mobile computing | Proof of Participation | Decentralisation through mobile computing with architectural entities of creator, validator, and constructor. | Only prototype has been built at this stage. |

## 2.2 Layer-One.S solution

In this section, decentralised scaling is introduced as a Layer-One.S solution. The component of decentralisation is to allow the participating nodes to be able to store, validate, and add transactions in the form of blocks on the chain. In the process of carrying out this procedure, the nodes need to perform computational work that puts a factor on time and hence limits scalability as the factor of scalability depends on transferring transactional value with optimal friction on the network.

<u>Layer-One.S components</u>
Generally, the concept of transactions per seconds depends on the following three factors:

1. Block Size.
2. Computational need or complexity to solve the block puzzle.
3. Relay time which affects the block finality and provides security.

In existing layer one solutions, the solution for scalability is based on one or more of the three factors where they have provided alternate solutions to the problems while not directly addressing that problem. Layer-One.S scaling solution is going to address the three factors above by using two important procedures.

The ability to map and store information respectively up to and on to the computational model for the mobile computing devices allows for cache threading and execution on a real time basis, whilst the decentralised application provides synchronous finality. This will be done using user authentication which allows for storage offloading and "Store-Based" validation. The abstraction level provided will not interfere with the operating system (OS) that the application is built upon, rather it will allow for transaction sharing and relay with the OS and the abstraction layer.

Layer-One.S functionalities are as follows:

1. Mobile Computational Dependency.
2. Computational Pool Sharing, using the ability to use computational based random delivery networks.
3. Roll ups using computational groups and full node threading.

Layer-One.S solution consists of the following three components:

1. Creators
2. Validators
3. Constructors

*Creators*
The creators are the transaction introducers. The ability to send transactions or state change functions through the abstraction layer is introduced. The role of the creators is to send the transaction into the pool of transactions that are held in a queue (negligible time) after which the transactions are then divided into and shared with the Directed Acyclic Graphs (DAG). The use of the DAG is done to maintain the level of growing concern amongst the nodes. DAG and blockchain are conjoined to provide a solution to the decentralised ledger technology.

Provisioning of security is provided by mixing the transaction pool into the DAG whilst maintaining a global state amongst the state channels using segmentation and division of information. The use of the DAG is primarily to get rid of the roll ups that are needed in the Merkle Tree which needs asynchronous reliability.

*Validators*

In a semi-decentralised network such as Ethereum and Polkadot, technologies are built to provide solutions complementary to each other however they still maintain a level of centralisation through division of transaction finality. For example, in Polkadot, components consist of collator, fisherman, nominator and validator. The validator has the ultimate authority being selected by the collators who stake for voting. This causes issues with the sharing of information, reliability on validators to maintain full nodes and a centralised architecture of para-threading/executing.

Introducing the dependency on mobile computation and the ability to execute computational pool sharing to provide random delivery networks for executing transactions from genesis until finality will help to offload the burden on full nodes and introduce para-data nodes where the nodes will store information and validate based on random data storage sequencing.

Once the constructors provide the transactions into the DAG, the validators from a pool of *n* form a consensus mechanism based on proof of participation. It allows for data validation through segmenting which allows for nodes to validate the piece of information that it holds.

This helps to validate and store information onto the blockchain in a completely decentralised manner and in the same instance, it allows for instant finality. This results in a completely decentralised network with the participating nodes whilst eliminating the need to maintain a full node.

*Constructors*

The role and responsibility of the constructors is state mapping into the DAG channel whilst updating the different states which are namely transactional canonical, date stamps, updating routes for information and the like. The constructors will allow for abstraction level mapping which is based on information grouping through the provisional DAG information. This will provide "Transaction Finality" whereby final states can be retrieved by the validators as the progress for finality is considered. Block finality and decentralisation are directly pegged to transaction validity in terms of maintaining a global state and in terms of information sharing across nodes.

<u>Transaction flow</u>

Initiating a transaction on the blockchain requires the initiator of the transaction to sign and send the transactional information to the broadcasting nodes after which the transaction status is sent back to the client side. Every transaction, which connects to the blockchain using a custom WebJS or through a third-party plugin, is looking to change the state on the blockchain.

# TRANSACTION INITIATION

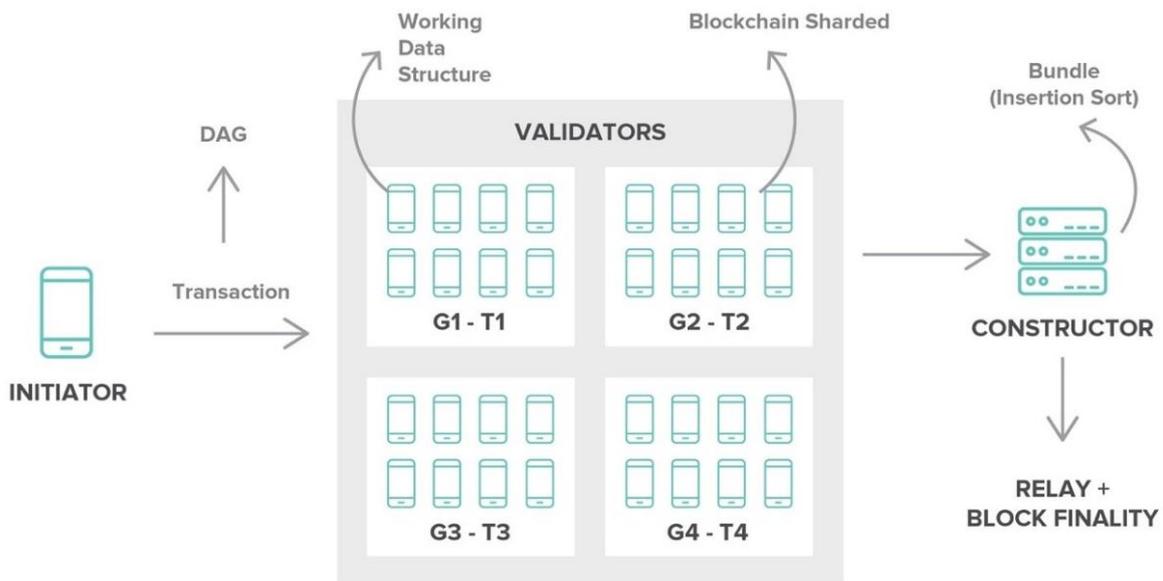

Figure 1. Transaction flow

*Step One: Transaction Initiated*
The initiator wants to send the transactional information to the network to change the state of the blockchain which can be sending a known value or executing a transaction with a binary request that facilitates an instant response. The initiator will send the request using Layer One library with set parameters in the object that is created. The request in the parameters and the information stored in the transaction being sent are (i) Node Groups, (ii) To, (iii) Value, (iv) Cert_ID, and (v) Hash_Data.

- Node Groups: The nodes are grouped into different tries in action which have the data and/or the data certificate for the validation. The grouping is arranged as follows: group 1 is for processing smart contract accounts, group 2 for transactions, group 3 for receipts, and group 4 for others. As shown in Figure 1, the nodes are randomly grouped into G1, G2, G3, and G4 for validating transactions i.e. T1, T2, T3, and T4 respectively. Note that nodes can be computers and/or mobile phones participating in the network.
- To: This represents the address of the contact or the smart contract that must be activated on the blockchain.
- Value: This refers to the values to be used to send it. The value will always be determined in the native Layer One tokens.
- Cert_ID: This will be paired with the certificate ID that is used to represent that linked format for the next set of information that needs to be stored in the database.
- Hash_Data: This is the hash of the function itself which acts the pointer.

Below shows a pseudocode of function to create transaction:

```
const initTxn = function create_transaction {
        const n_g = new node_group (get_node_status)
        to:
        value:
        cert_ID:
        hash_data (this.hashit)
} // Not using await. Promises are not needed until the second stage.
```

*Step Two: Certification Issuance and DAG activation*
Every transaction going into the network is issued with negatively correlating (complexity, resource and time) task intensive custom hashes which will act as the identifier and the pointer for that particular transaction, called a certificate. The hash is then added into a certificate which solves two purposes i.e. (i) indexing validation and (ii) roll-ups.

The transactions are introduced and processed on the network through a Directed Acyclic Graphical (DAG) process which is blockless in nature. DAG provides a topology which will be applicable for the blockchain based trie structure (digital tree) for the node selection. The transactions are a part of the vertices, and the edges form a part of the direction to the nodes that are grouped on the network according to the sharded ledger that they maintain, they are randomised frequently.

The transactions can be authenticated and passed to the nodes in sequence and due to the higher throughput nature of the consensus mechanism that proof of participation activates for the network; parallelism is achieved from the nodes and the DAG structure. Every transaction is initiation with a cycle counter. The cycle counter is used to synchronise DAG and blocks.

*Step Three: Validator Indexing and Assignment*
The transactions are cycled and forwarded to the network nodes where the nodes are grouped according to the sharded ledger that they maintain. The groups are created and based on tries.

Once the nodes are broadcasted; the consensus mechanism calls the validation function to be executed and the nodes are selected through a *custom random node selector algorithm (node algorithm)*. The random node selector forms groups on every *n* cycle and validates the transactions. The validation is dependent on the node holding the information and certificate that go to validating the transaction.

Layer-One adopts the process of Sharding and parallel processing to perform such actions. To store information on the nodes and/or certificates which access the information stored on an optionally opted information warehouse for private chain transactions.

Information is stored through horizontal Sharding by creating a custom key to hashed pointer based Sharding and adopting ranging of information to avoid hotspots. Randomly redistributing information on the nodes through node algorithm prevents bad actors on the network taking advantage of resource aggregation.

To overcome the unbalanced distribution of information storing and validation, the cycle of information distributes with a randomly selected fixed size (tries) of information that is stored based on *hashed characteristics*.

*Step Four: Construction, Relay and Block Finalisation*
After the nodes have finalised the transaction and updated the state; the constructor saves the image of the state into a certificate and makes sure that the blocks are relayed and added back into the network while ensuring that the state has been saved.

The constructor is a randomly selected group of nodes / system coming together to save the state and issue a certificate which is based on the validator certificates. This is the third state certificate; the first and the second being the initiator and the validator certificates respectively.

The information is relayed back into the network and the block is finalised whilst updating the nodes. This process allows for the state of finality to be certified. The process of maintaining key-certificates and the data is based on a custom Markel Tree and tries.

# 3. Interoperability

Blockchain interoperability refers to a composition of homogeneous and heterogeneous blockchains among which a source blockchain is able to change the state of a target blockchain, empowered by cross chain or cross blockchain communications and/or transactions.

## 3.1 Review of interoperability limitations

The landscape of blockchain interoperability and existing technology solutions has evolved over time. We classify the blockchain interoperability solutions into 3 categories i.e., the first, second, and third generations of blockchain interoperability solutions. Layer-One.I is the blockchain interoperability solution that will enable data and value transfer across different blockchains. Table 2 presents the existing interoperability solutions, against Layer-One.I solution.

The first generation of blockchain interoperability solutions identifies and describes different interoperability strategies across blockchains including sidechain and relay approaches, notary schemes, and hash time lock contracts. When a blockchain considered as a mainchain has another blockchain as an extension of itself, the child chain or the extended chain is the sidechain. The mainchain maintains a ledger and connects to the sidechain via a communication protocol that facilitates asset transfer between the mainchain and the sidechain [47]. A sidechain can offload transactions from the mainchain by processing transactions within the sidechain and then direct the processing outcomes back to the mainchain. Mainchain may be a sidechain of another mainchain [48]. However, the exchange of messages is tied vertically, and it does not allow communication across or between sidechains. This is the case when transactions need updating or syncing simultaneously. For example, in payment channels [40], the cryptographically signed messages update the current state without broadcasting it to the mainchain. Only when the payment channel is closed, is the final state published onto the mainchain. This may result in an inconsistent state occurring at the closure phase [49].

A relay approach requires the presence of a sidechain. In other words, there are no relay solutions without a sidechain [6]. In relay solutions, relayers keep track of the block headers of the mainchain, a blockchain network e.g., Bitcoin, and input them to the relay smart contract hosted on another blockchain e.g., Ethereum. A pool of block headers can be used for verification of transactions or information [50]. Security limitation is inevitable as we have to assume that the mainchain is secure. Another limitation is around centralisation in sidechains. It is likely that there will be fewer validating nodes in the sidechain to compromise with high throughput. Bad actors can then take control of a potentially small set of validators. Conversely a strong security model in a sidechain is generally a trade-off for a slow transaction settlement, stalling assets, and lowering liquidity [6], [51].

Secondly, Notary schemes involve an entity that monitors multiple chains and triggers transactions in a chain through deployment of a smart contract upon an event taking place on another chain [48]. Notary schemes are practically instantiated as cryptocurrency exchanges, facilitating indications between market participants by managing an order book and matching buyers and sellers [6] for example Binance, Coinbase, etc. Notary schemes are fundamentally intermediaries between blockchains as they execute actions on behalf of the end-users or facilitate matching for the end-users with trade offers [52].

Thirdly, Hashed Time-Lock Contracts (HTLCs) enable atomic operations [53] between different blockchains using hashlocks [54] and timelocks [55] allowing asset exchange without a pre-existing trust relationship between the two different blockchains. Even though HTLC concepts of hashlocks and timelocks exist at Layer two, the protocol and governance in HTLC implementation are different across existing solutions which result in silos. Generally, HTLC technique is similar to programmable escrows hence it is practical in a trustless system and without a third party, unlike in the notary schemes.

The second generation of blockchain interoperability solutions provides the ability to create application specific blockchains that can interoperate between the customised blockchains. This generation of blockchain interoperability solutions use strategies of bridge or hub-and-spoke. Polkadot [56], [57], Cosmo [58], Ark [59], Komodo [60], and AION [61] are examples of this generation. Foundation that Polkadot provides is Parachains, which are the parallelised chains that participate in the blockchain networks. Parachains use bridges linking independent chains [57]. Polkadot achieve interoperability through the chain-relay validators to validate state transition. Parachains communicate through a queuing mechanism based on a Merkle tree [62]. Polkadot scales by connecting more parachains to the relay chain. Cosmo is a decentralised network of independent parallel blockchains called zones [58]. That said the zones are basically Tendermint blockchains [63]. Data can be transferred between zones via hubs [6]. Double spending can be avoided by minimising the number of connections (via hubs).

The third generation of blockchain interoperability solutions known as hybrid connectors attempt to deliver a blockchain abstraction layer in which a set of uniform operations can be exposed to allow an interaction between blockchains without the need of using different APIs [64]. Solutions such as Trust relays, Agnostic Protocols, and Blockchain Migrators are in this family. For trusted relay solutions, trusted escrow parties route cross blockchain transactions from a source to a target blockchain, allowing end users to define their own business logic [6]. As such, different APIs have to be managed and cross chain consensus has to be implemented as well. Trusted relay enables interoperability through a set of interoperability validators who are participants from the source and target blockchains. Cross chain validators collect transaction requests, validate, sign, and deliver. Agnostic protocol enables interoperability by providing a blockchain abstraction layer [64]. For this solution, a consensus protocol organises blocks that contain a set of transactions spread across blockchains giving a holistic updated view of each underlying blockchain [6]. Blockchain migrators allow an end-user to migrate data and possibly smart contracts across blockchains [65]. An abstraction of smart contracts can be sent across blockchains allowing the state changes required to be executed [66].

Table 2. Existing interoperability solutions, against Layer-One.I solution (adapted from [6])

| Interoperability project / Blockchain | Main Chain | Consensus Protocol | Description | Limitations |
|---|---|---|---|---|
| **BTC Relay** [67] | Ethereum | Proof of Work | Simplified Payment Verification (SPV) approach relying on block header verification. Ethereum smart contract reading Bitcoin's blockchain. | Limited functionality. |
| **Peace Relay** [68] | Ethereum | Proof of Work | SPV on Ethereum Virtual Machine (EVM) based blockchains allowing for cross chain communication using relay contracts. | It is computationally expensive to verify block headers. |
| **Testimonium** [69] | Ethereum | Merkle proof of membership | SVPs relying on meta data block header verification (e.g., block headers store the block number no transaction data). | Mainly support Ethereum virtual machine based blockchains. |
| **POA network** [70] | Ethereum | Proof of Authority | Applicational interoperability to EVM-based dApps. | Validators confined to a geography. |
| **Liquid** [71] | Bitcoin | Strong federations | Strong federation of functionaries maintain the settlement network. | Consensus secured by specialised hardware. |
| **Loom Network** [72] | Ethereum | Delegated proof of stake | dApp platform with interoperability capabilities to interconnect with other major blockchains (Ethereum, Binance Chain, and TRON networks). | Not an open-source solution. |
| **Zendoo** [73] | Bitcoin | Proof of stake | zk-Snark solution allows the mainchain to verify the sidechain without disclosing sensitive information through sidechain creation platform. | Zk-Snarks are computationally expensive. |
| **RSK** [51] | Bitcoin | DÉCOR+ | Federated sidechain, in which RBTC (native currency) is tethered to BTC. | Relies on selfish-mining, energetically inefficient. |
| **Blocknet** [74] | Ethereum | Proof of stake | EVM-based blockchain allowing trustless blockchain interoperability. | Limited functionality. |

| Interoperability project / Blockchain | Main Chain | Consensus Protocol | Description | Limitations |
|---|---|---|---|---|
| **Wanchain** [75] | Ethereum | Galaxy protocol | Cross-Chain bridge node staking rewards with following entities: vouchers, validators, and storemen. | Storemen are not truly decentralized. |
| **Lightning Network** [40] | Bitcoin | Proof of Work | Relies on multi-signature channel addresses enabling high volume, low latency and micropayment. | Timelock expiration exploits. |
| **Tokrex** [76] | (not defined) | Proof of Work | Validation and escrow nodes distribute key generation enabling cryptocurrency exchange. | Security is compromised because the private key used for asset transfer are known by both sender and receiver know. |
| **Fusion** [77] | Ethereum | Practical Byzantine fault tolerance | Distributed storage of a private key and crypto asset mapping. Distributes trust and responsibility of managing private keys. | Does not provide instant atomic swaps. |
| **Xclaim** [78] | Bitcoin, Ethereum | Proof-or-punishment | HTLC-based trustless protocol that manages cryptocurrency-backed assets with following architectural entities of Requester, sender, receiver, redeemer, the backing vault, and issuing smart contract | Could lead to locked funds. |
| **DeXTT** [79] | Ethereum | Proof of Intent | A protocol implementing eventual consistency for cross-blockchain token transfers. | Strict requirements of PoIs. |
| **Xchain** [80] | Ethereum | Avalanche (optimized DAG) | Generates custom smart contracts for performing cross atomic swaps. | Only applicable to Ethereum. |
| **Cosmos** [58] | Cosmo hub | Tendermint | Each parallel blockchain or zone is self-sovereign hence there is no limit for how many zones can be attached to a hub. | Rely on a bridge-hub architecture hence do not support horizontal scalability through sharding. |

| Interoperability project / Blockchain | Main Chain | Consensus Protocol | Description | Limitations |
|---|---|---|---|---|
| **Polkadot** [56] | Relay chain | Grandpa and Babe | Relay chain connects between para chains and para chains to bridges. Architectural entities handling transactions include collator, validator, nominator, and fisherman. | Rely on bridges that route transaction from a particular blockchain to another type. |
| **Ethereum 2.0** [81] | Beacon chain | Serenity | Composing with shards that interoperate with each other. | Rely on a sharding strategy. |
| **AION** [82] | AION chain | Proof of intelligence | Using bridges to connect and route patriating networks with connecting networks. | Highly experimental products. |
| **Komodo** [60] | Bitcoin, Ethereum | Delayed Proof of Work | Provides a decentralised exchange framework for cross-chain atomic swaps with following entities: liquidity provider nodes, buyers, and sellers. | Highly experimental products. |
| **Ark** [59] | Ark chain | Delegated proof of stake | Smart bridges making instances of its platform interoperable. | Not entirely decentralised as intermediary nodes are necessary to achieve interoperability on an ad hoc basis. |
| **Trust relays** [83] | Hyperledger | Validator quorum | trusted escrow parties route cross-blockchain transactions | Not entirely decentralised as blockchains participating in the network know how to address each other. |
| **Interledger protocol** [84] | Private, Public | Packet switching | Packet switching (ILPv4) defines how senders, routers (or node, or connector), and receivers interact. | The root of trust is the connector which has to be trusted. |
| **Hyperledger Quilt** [85] | Private, Public | Packet switching | Quilt implements several primitives of the Interledger protocol. | It does not support the transfer of non-fungible tokens. |

| Interoperability project / Blockchain | Main Chain | Consensus Protocol | Description | Limitations |
|---|---|---|---|---|
| **LayerOne.I** | LayerOne chain | Proof of Participant | Interoperability through true decentralisation with flash contracts and nucleus scripting. Architectural entities include initiate, lock, flash, sync, and publish. | Only prototype has been built at this stage although the development of cross blockchain transfer has been finalised. |

## 3.2 Layer-One.I solution

The Layer-One.I framework allows for exchanges between various value-based parameters such as assets and transactions, linking chains based on information that can be provided between cross blockchain transactions. The interoperability includes innovativeness in value exchange by implementing flash contracts, nucleus scripting, data synchronisation, side chain and locker technologies.

The steps for cross blockchain communication are as follows and also shown in Figure 2.

Step 1: Identity management
The cross blockchain communication layer allows for identifying the sender and receiver in the transaction by the "Layer-One.W" which is the authentication mechanism to sign and broadcast the transaction.

Step 2: Flash contract
Flash contracts are the process of automating the decentralised pool allocation through identifying the participants and making the exchange happen in between them. Flash contracts allow for the code to be executed on the nodes by pooling a distributed set of a *required resource*.

The asset of A is locked into custody on the network allowing for the script to find the value acceptor. Once the value acceptor has their value checked, both the assets are locked into the pool ready to be exchanged. This process allows for the global state of the blockchain to be maintained.

Step 3: Settlement through scalability and pool using customised Directed Acyclic Graph
While the process of getting information from the decentralised pool is maintained, the null pointers are maintained by a checksum strategy that will allow for minimal resources to be checked which allows it to grow exponentially instead of linear growth passing the scalability test of the Big(O) notation.

Step 4: Nucleus Scripting and asset locking
Once the state has been changed in the network, there is an exchange that takes place which allows for nucleus scripting to run through sharding and sending the updated global state to the network. The process of changing the global state is achieved by using custom data structures.

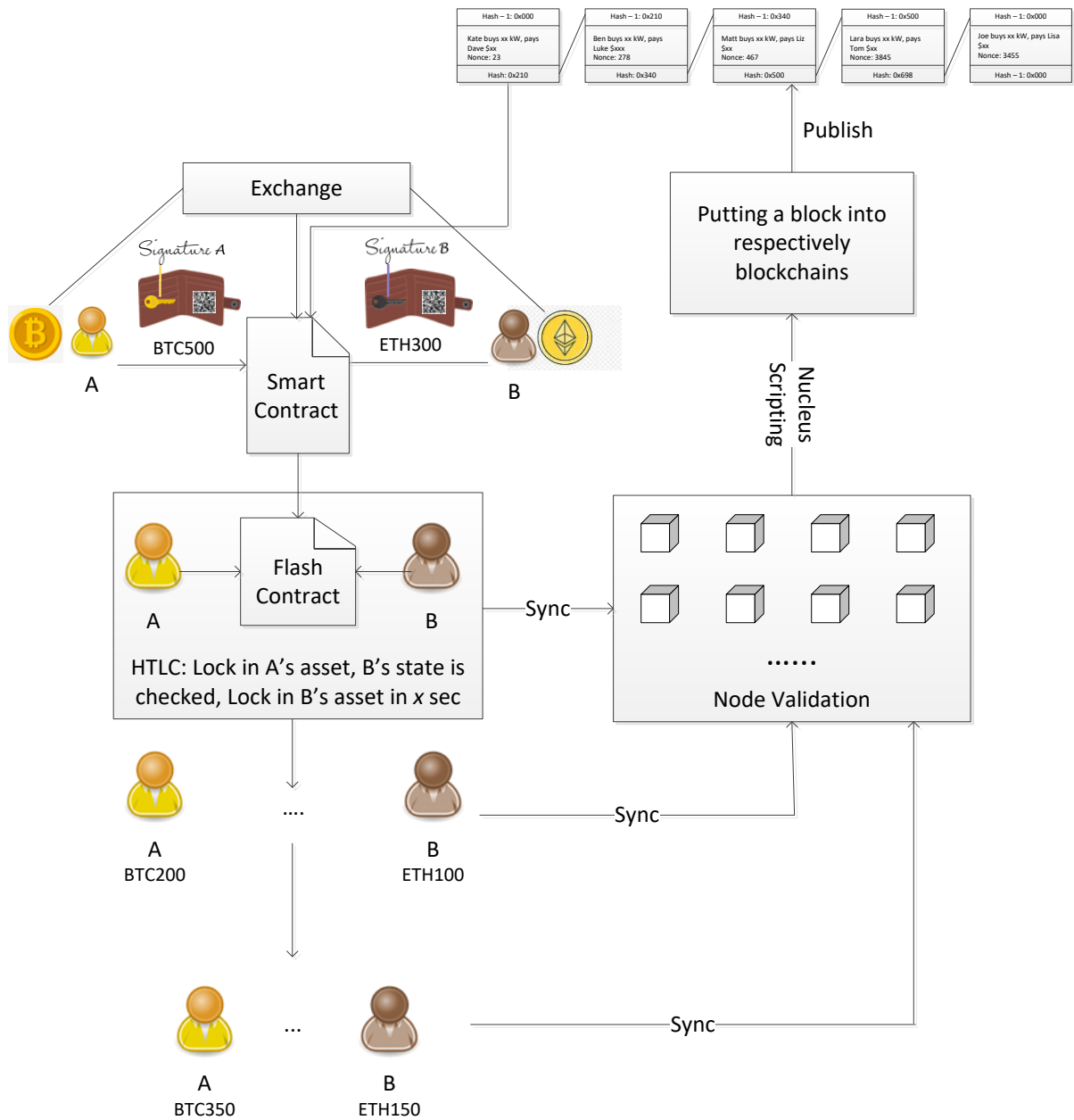

Figure 2. Exchange across blockchains over the Layer-One.I framework

## 4. Mobile Phone and Nodes

Nodes act as the DNA in the blockchain as they identify the processes, validate, and append information in the form of blocks onto the chain. The method to validate transactions and add them into the block to achieve finality requires mining or staking which relies on a cryptographic protocol and the ability to reach a consensus, for example, closest checksum to a hash.

There is an inherent limitation in this approach which is that the algorithm scales linearly in correlation to the resource utilisation. The $t$ to maintain the velocity in the supply side has to be frequent in order to maintain an optimal level of friction on the network, as this will allow for exponential scaling on the Big(O).

There have been changes implemented overtime through soft forking to optimise the code, to execute transactions based on binary and to customise linked list structures. The fundamental problem of reaching optimum Tera hashes per second (TH/s) can only be achieved with optimal hardware. To overcome the problems, the use of mobile enabled computing and validation will provide a node and data store facility to pool computing and data storage. This introduces a scalable, secure, and interoperable consensus mechanism which is called "Proof-of-Participation".

### 4.1 Distributed node and storage

The procurement of mobile-enabled-devices (MED) for validation, publication to the network (validation), storage, and operation of scripts will enable a plethora of opportunities to distribute transactional capacity for true decentralisation on the network. The transaction operation and validation process will be executed on the node using scripts running on a MED and will be customised as per the floating-point operations with the need to make the authentication process optimal for the task and the resource.

The role of mobile based computing allows for the limitations of the current validation, mining and staking problem such as the process of computing nonces to solve the problem of leading zeroes. Mobile computing will be able to introduce systems such as independent submissions for the device actions such as identification for access modules, gateway channels and target rendering services. The distribution of the workload via P2P will allow for the computation to be completely decentralised rather than maintaining nodes of information that is centralised since the genesis block has been generated.

The problem with micro transactions or payments with the scripts running on the nodes is that it consumes the resources on the network, increases the cost of such transactions and does not allow for an economically feasible solution to scale for mass adoption. The proposed solutions with mobile computing will allow the developers to access the network with an abstraction level of assistance which will be coupled with a loose governed set of API's and index level SDK's.

### 4.2 Innovating the process of executing scripts and thus transactions

To overcome the problems that mining enforces on the data structure appending; the consensus for Proof of Participation enforces using DAG to add and traverse the transactions in a computation-less manner until Nucleus scripting enforces cryptographically adding blocks by sharing computation resources that can be leveraged from the mobile nodes on the network. The transactions enter the acyclic graph via the initiator whose identity is protected by a custom Private-Key-Generator which allows defence against identity and spoofing.

The ability to use mobile computing to verify transactions through sharding of threads enables the developers to cope with the abstraction features by not only making sure that the hardware of the devices is taken into account on a self-learning basis but also the transaction and execution split can allow for retrieving information directly.

This will lower the redundant puzzle solving and allow for higher throughputs and thus reducing the cost of utilising the blockchain network. The mix of DAG and blockchain to innovate the concept of resource sharing can be utilised in various industries for application building such as data validation, payments including micro sum transactions, tokenisation of web, products and others.

### 4.3 Data structure using Directed Acyclic Graph and Blockchain with modified Merkle Tree

To allow global adoption for Decentralised Ledger Technology (DLT) and to have transactions, data storage and access, validation, and process on the DLT, there have been various suggestive innovations such as increasing block size, reducing processing time or puzzle complexity on the network. One of the innovative concepts has been DAG.

DAG allows the blockchain to scale as they have an efficient transaction sequencing structure that allows the DLT to store and retrieve information quickly. Until now, DAG has been used in isolation to blockchain. An analogy to this is that an insert algorithm is faster than a quick sort whereby DAG is like a quick/merge algorithm. Insert and sort must work with the quick/merge algorithm to achieve the desired result. Hence, we make use of DAG paired with Blockchain along with Proof of Participation.

IOTA and ByteBall have used DAG in their technologies to scale but the problems of centralisation and bad actor takeover have remained. With Layer-One solution, we will solve both the problems that IOTA and ByteBall have experienced. Ethereum makes use of different tries to be able to store, process, parse and validate information on the blockchain. Layer-One solution will be making use of DAG to complete the transactions and then store the information through a custom zk-Snark technology in the form of blocks. This allows for faster transaction throughput and information storing on the network.

There will be four structures of DAG maintained which will allow for a global state, transaction and information state, storage state and parsing stage which will index the information and keep a record of the latest state changes. These four stages will work together to be able to store, validate, transact, and prove the transactions that are done on the network. The role of zk-Snarks implementation in the parsing stage will allow for minimum resources to be spent on Big(0) n*n matrix where the transaction affirmation will be done on a quicker basis which will also provide an abstraction layer in order to complete the work.

## 5. Tokenised Economy

Blockchain's revolutionary decentralised model has seen increased in financial sectors worldwide and monetary features have been the most explored and applied in blockchain project [86]. Groups of producers and groups of consumers have been connected to create transactions, with tokenisation being used to incentivise producers to participate through collaborative value generation on the network.

Transactions can represent any form of value including monetary currency, physical assets, intangible assets, access to a specific utility, etc. [87]. Crypto tokens are fundamental to the value transfer mechanism [88] and are designed to be unique, liquid, secure, tradable, transferable and digitally scarce [89]. Crypto tokens can be defined as digital assets that use cryptographic techniques to regulate the generation of units of the asset and to verify their transfer between parties via a blockchain [90]. Blockchain networks are ideal infrastructures to build token economies in a P2P network without centralised intermediaries such as banks or large enterprises. Over 6,050 different crypto tokens exist at the time of writing (Aug 2021) [91]; Figure 3 shows the growing number of cryptocurrencies from 2013 to 2021.

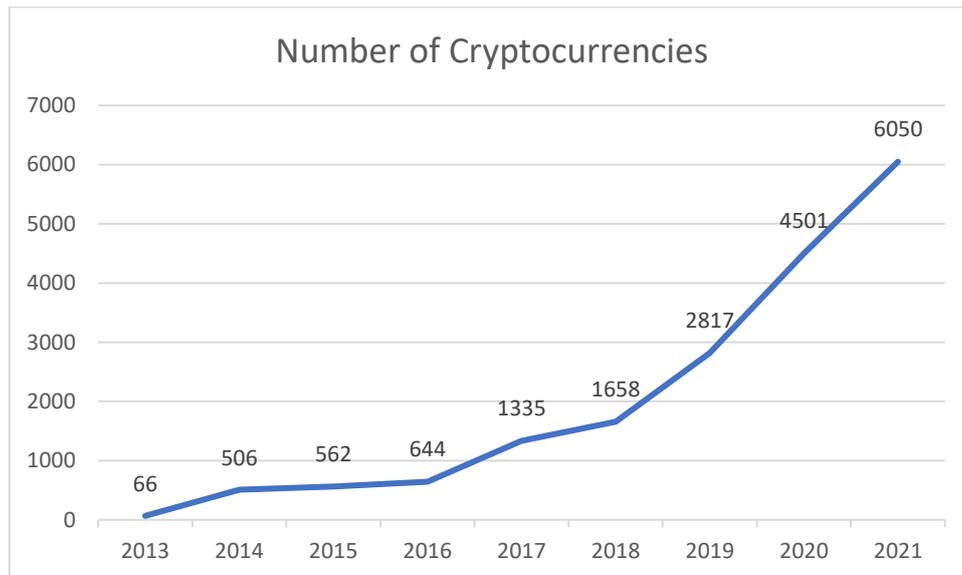

Figure 3. Incremental number of cryptocurrencies between 2013 and Aug 2021

## 5.1 Review of token economies

In this paper, utility tokens are explored. There are five models of utility tokens exist based on the nature of the work being provided to the network i.e., work, service, governance, discount, and burn and mint tokens. The five types are outlined briefly as follows.

In the work token model, a service provider stakes the native token of the network or dApp to earn the right to validate work for the network [87]. This model also works like taxi medallions in which service providers must pay for the right to work on the network [92]. The work tokens can be used to bootstrap and coordinate the supply-side of a network in a way that would otherwise be very difficult and even impossible, by providing an incentive in the form of potential yield for those staking to provide services to the network. In some cases, work tokens serve as a mechanism design tool enabling 'skin in the game' for service providers. That says users are not only rewarded for good work but can also be punished for the work that harms the network without relying on identity and/or reputation. The ICO can focus on the sale of the validator token. The probability that the service provider is awarded the next job is proportional to the number of tokens staked [93]. A variant of work tokens is to create tokens, one to pay for underlying goods and the other that validators must stake to obtain the right to record the next block [94]. The reward that the validator gets is paid with the token that can be used to buy the underlying goods. Examples of token utilised the work token model include Augur (prediction markets) [95], Keep (off-chain private computation) [96], Filecoin (distributed file storage), Livepeer (distributed video encoding) [97], Truebit (off-chain verifiable computation) [98], and Gems (decentralized mechanical Turk) [99].

In the service token model, a user stakes tokens in order to provide a given service to a network in exchange for cash flows. If the work is done correctly, the user receives a reward comprising of either fee paid by the demand side or inflation. Crucially, these fees need not be paid in the native token as long as the native token must be staked in order to qualify to receive fees. However, if work is done incorrectly or maliciously, the stake can then be slashed. Different implementations of service token designs include (i) 'Skin in the game' tokens, where the amount of tokens a user has staked influences the amount they can earn; (ii) 'Token Curated Registries', where users can stake tokens in order to filter data in and create trusted lists; and (iii) 'Burn & Mint Equilibrium', where, rather than paying fees to service providers, users burn tokens in the name of a service provider and service providers receive a pro-rata share of monthly inflation based on percentage of tokens burned in their name.

In the governance token model, the governance of a blockchain network is decentralised to empower the community to participate in a democratic manner. Holders of governance tokens have power to influence decisions on the core protocol, the evolution or upgrade of the network, product and roadmap, hiring, token treasury management, and changes to governance parameters [100]. A governance token gives users the ability to influence the way the network is run, including anything from electing representatives, proposing and voting on network upgrades, deciding how funds are spent and even determining monetary policy. While almost all cryptocurrencies possess some level of governance through off-chain communication as well as the ever-present possibility of forking the protocol, this kind of informal governance is generally referred to as "off-chain" governance, whereas governance tokens refer to projects implementing formalized, automated systems of "on-chain" governance in which governance rules are encoded onto the protocol and the results of the governance process are automatically executed. The intuition behind governance token valuations is that as the value of a network goes up, the ability to influence how it is run should become a scarce resource, with some, like Fabric Ventures, even arguing the value of this influence will actually scale exponentially with the value it secures. One of the well-known examples of this token model is MakerDAO (powering the DAI stablecoin) [101].

In the discount token model, discount token holders are entitled to a perpetual discount on services offered through the network or dApp. Nevertheless, the discount is mathematically equivalent to a revenue or fee royalty, given that the token holders will be qualified for the benefit only if they use the services. The discount token holder is designed to grow in line with the utilisation of the network-based services [102]. Variations of this model are around the holding or staking tokens for membership access and using or destroying the tokens to obtain discounts [87]. Examples of token utilising the discount token model include Binance's BNB (use discount) [103], Sweetbridge (discount) [104], and COINFI (membership access) [105].

At a high level, a discount token grants the user discounts on transactions performed on the network. Discount tokens can be implemented as a "Use" model in which case a discount is provided for users paying for the service using the discount token or as a "Stake" model in which users must stake the discount token in order to qualify for a pro-rata share of the total available discount. The SweetBridge "Stake" model leads to far higher value capture than the "Use" one, as in the latter the discount token is used as a medium of exchange to pay for discounted fees and thus suffers from the velocity problem. The "Stake" discount token model possesses several interesting features since the discount provided by each token is mathematically equivalent to a cash flow, it can be valued using a discounted cash flow. As a result, the token's value is directly linked to transaction volume and grows alongside network adoption. Despite the fact it can be valued similarly to an equity or other cash flow generating instrument, it necessarily qualifies as a utility token since the value flows are only received if the user utilizes the platform. The token does not interfere with the UX as users do not have to transact in it. The tokens are actually beneficial to users over speculators.

These are arguably a type of service token but one in which, rather than paying fees to service providers, users burn tokens referred to 'Burn & Mint Equilibrium' model. In this model, users do not actually have to stake the native token in order to receive value flows. Instead, the token acts as a proprietary payment currency with value flows being passed on indirectly through deflation.

In this model, proprietary payment and tradable tokens are utilised in a two-token system. These two tokens can be exchanged at a fixed rate designated by a fiat currency. In order to use the service, users have to burn the value-seeking token to receive the payment token [106]. The payment tokens are minted by burning the value-seeking token. Examples of tokens utilising the burn and mint token model include Factom (data integrity) [107], Scriptarnica (e-book reading, writing, publishing, buying and selling) [108], Gnosis/SpankchianCivic (adult entertainment videos) [109], Golem (computer power rental) [110], Basic Attention Token (content creators and publishers) [111], 0x (decentralised exchange) [112], Helium (decentralised wireless network) [113].

## 5.2 Layer-One.X tokenomics discussion

Blockchain-based token economy design requires a rigorous engineering approach to be able to sustain and lead to a virtuous cycle of growth driven by network effects. Token engineering can be defined as the theory, practice, and tools to analyse, design, and verify tokenised ecosystems [114]. Token engineering encompasses theoretical concepts being applied in order to create and deploy a token economy. Defining the main goals of the token system is a critical step in token engineering leading to rigorous analysis, design, and verification of ecosystems and desired behaviours.

Layer-One.Coin (Layer-One.C) measures its stability and facilitation of transactions by engaging its resource accessibility model to tend towards the right balance between the supply and demand intersection to be determined by time $t$ and resources $r$ on the network.

The goal to reduce the friction to an optimal level based on the correlation in sequence of:

- Decrease in friction/transaction cost leads to,
- Increase in token transaction which leads to,
- Increase in token velocity which leads to,
- Increase in Supply of tokens which leads to,
- Decrease in the Dollar Value of the token.

The factor of $t$ and $r$ on the network are directly correlated to Demand $D$ and Supply $S$ on the network. Here we correlate the friction of the token with the number of Groups and the Consensus Mechanism of relatable transactions.

Friction on the network
Friction $F$ on the network determines the excess the transaction initiator is willing to bear to conduct a transaction on the network. The goal of the $F$ is to be at optimal. Three situations emerge out of Friction as follows.

*When F < 0 (not ideally possible)*

Where the number of transactions must be recorded on the network in a time interval, the number of transactions should not have latency time and cost associated. Where $F < 0$, validation compensation will be less than zero which is not economically feasible as there is a positive correlation in between validation compensation and the time required to record a transaction.

If the network uses its resources to conduct transactions, the market cap of the total volume will go down which will drain resources on the network. This will not provide a scalable mechanism of conducting transactions.

*When F = 0 (no growth in the token)*

This will lead to no growth in the value of the tokens as the friction cost will increase on a constant basis if the supply and demand increase on a constant basis. If there is an imbalance in between the supply shift, the impact will be on the friction and hence this model is not suitable for an imperfect marketplace for transactions.

*When F > 0 (Inflation, Mint and Burn)*

Where the friction for $F > 0$, PoS and PoW based consensus mechanism will create ways for accumulation problem. Velocity will be lower, and this has a direct correlation on the transaction cost increasing.

When the friction is set by the network based on the Proof of Participation, it will adjust the friction curve based on the supply and demand of the transactions on the network. The goal is to keep the velocity of the token at an optimal network acceptance level.

$$F = f(S, D)$$

$$S = f_S(\sum_{i=1}^{4} n_{Gi}, Tr(t,r), v)$$

$$D = f_D(\sum_{i=1}^{4} n_{Gi}, Tr(t,r), v)$$

Where $F$: Friction
$S$: Supply
$D$: Demand
$n$ : Number of nodes
$G$ : Groups of (i) smart contract accounts, (ii) transactions, (iii) receipts, and (iv) others
$Tr$ : Transactions
$t$: Time
$r$: Resources
$v$ : Velocity

The friction is paid by the initiator and the allocation of the friction is put back into the pool which is distributed to adjust supply curve into the system. The more transactions that are processed on the network by an initiator, the more resources the initiator will pass on to the network to maintain the pool.

Consensus Mechanism and Incentive Scheme

The Proof of Participation consensus mechanism proposed on the system is based on a resource allocation module where the participants in transaction or value transfer constitute the nodes and the network. We will use user $u$ as the initiator and the nodes $n$ as the verifier on the network. Hence,

$$F = 0 \rightarrow V = \{G1, G2, G3, G4\}$$

Where $F$: Friction
$V$ : Validator
$G$ : Groups of (i) smart contract accounts, (ii) transactions, (iii) receipts, and (iv) others

Resource Costing Model

Cost of the resource allocated must be based on the equilibrium of initiation and validation. For the total number of transactions, there is an equal amount of demand based on the network. Hence, Supply and Demand will be 1:1 with an error of margin of 1(10%):1(10%) which will be corrected by the network.

Token Distribution and Pricing Strategy

The distribution of the Layer-One tokens is based on building network effects and bootstrapping the network from the initiation phase, including participants such as layer two application developers, network growth facilitators and users. A pricing strategy that is optimised for growth will be adopted. There are various correlations with the factors that comprehend the token distribution model.

*Consensus Mechanism and Token Pricing with respect to token inflation.*

Layer-One.X and its innovative Proof of Participation based consensus mechanism provides an acceptance state of validation. When the participants are validators, they are a part of the same pool of the network. The resource utilised to achieve a consensus state can be directed towards an optimal resource utilised and incentive mechanism. When the incentive mechanism is optimal, the velocity of the token for rewards is optimal.

If there is an imbalance in between the incentive and token velocity, the price of the token will incur more inflation. Layer-One.X has inflationary measures taken into account.

For instance, FLOW tokens (the native currency of the Flow network) distribute the inflation to stakers and this results in a decrease in the value of the token. Evidently, the control of inflation must be done at the consensus mechanism stage.

*Token participant distribution and its pricing*

The participant token distribution strategy incorporated at Layer-One.X provides for growth in the network and incentivizes participants to get on board with the mission.

At Layer-One.X, the resource model is based on the usage of the computing power that an account requests and provides on the network where the participants and the validators are mutually exclusive but form a part of the same pool. In such instances, the participants that provide resource utilisation and availability are the developers and the users on the network.

*Ongoing distribution rules, Governance, and Token Pricing*

The ongoing distribution falls under two major components and its relation to token pricing is also considered.

Firstly, the technical module considers the resources available in the network and the cost to facilitate state changes. The value of the resource is based on optimally setting the value of the token to avoid considerable friction that will bottleneck the network.

Secondly, network and service level issuance consider different components such as ongoing mainly ancillary allocation of resources such as threading, spills, bundles, and others.

*Layer two tokenomics VS Layer one tokenomics*

The distribution of tokens to the developers enables the distribution and utilisation of the resources on the network which enables ongoing token distribution to increase token decentralisation as the users which act as the validators get rewarded.

To overcome the problem of increasing the cost of resources on the network, resulting in token inflation, the ongoing token distribution model needs to correlate with the growth of the network.

*Note that detailed information on token distribution will be provided in a separate document and is available upon request.*

### HeartBit Coins

Layer-One coins, HeartBit, are required in the following cases:

- Resources: To conduct a transaction on the network or to be able to execute scripts on the network. Resources on the network are the primary driver and HeartBit coins are required.
- Exchange: HeartBit coin provides a medium of exchange amongst its participants on the network which can be used to transact on the network.
- Liquidity for secondary tokens: To gain liquidity on the platform, HeartBit coins will provide access to decentralised pools that will allow for token exchange. They will also be able to act as collateral for secondary tokens.
- Rules and governance participation: For the ongoing rules and governance changes, it is important for the network to be able to sustain changes and updates. To validate updates, HeartBit coin will allow for the users to participate in governance and rule updates.

## 6. Use Cases of Layer-One.X

Benefits of the blockchain interoperability and scalability can be witnessed in various industries such as healthcare, financial market and technology, logistics and supply chain management, data crowdsourcing and tokenisation and user management and validation among many others. Here in the following subsections are some interoperability and scalability use cases that will be built on a layer two with the Layer-One scaling and interoperable solutions.

### 6.1 Interoperability value to the internet and industries

#### Staking through decentralised pools

Layer-One.X solution will permit developers to use the abstraction layer that will enable the crypto holders such as Ethereum to make use of decentralised pools. It will allow for cross cryptocurrency staking, reducing the staking cost through centralised exchanges that are currently charging over 10-30%.

Exchange of tokenised values

The exchange of tokenised values will permit the industrial use case to transfer the tokenised assets cross blockchain which will permit value, data, and transactional exchange.

Whilst there are a few solutions that allows for heterogeneous blockchain transactions (e.g. Wanchain), it converts the native cryptocurrency into a subset token called WanX (for example WanETH) which increases the transaction cost and time to exchange values. Layer-One.I will focus on decentralised pool exchange using the wallet integrated mechanism that will allow the users to authenticate transactions without transferring and changing the native global state.

Ledger tokenisation

With the advocation of Decentralised Ledger systems and its access to be able to securely store confidential information and share it amongst third parties with a commercial standpoint, it is imperative to establish a standardised ledger tokenisation system. Layer-One.X provides the same with use case in Intellectual Property (IP) and its functionality such as (i) tokenising information and agreements through smart contracts, (ii) proof of ownership and version control, and (iii) unifying the IP system.

Regarding IP ownership rights and sharing the rights with other parties, the modules or the rights to use certain parts of it can be tokenised using Layer-One.X smart contracts. This will allow for the information to be shared securely with resource and time factor attached to it. The tokenisation process can be applied towards the complete spectrum of parties such as suppliers, distributors and manufacturers that will permit for data tokenisation and commercialisation.

IP ownership can be securely communicated into a hash table with its version control. The different versions can be linked allowing for the data to be sharded and stored securely on the nodes on the blockchain. Layer-One.X will provide use case for IP tokenisation with different version control system that will allow the data to be hashed and shared with security measures. Confidential ledger is maintained and shared with parties on custom terms and conditions allowance for trading on the marketplace.

With the interoperable nature of the Layer-One.I technology, it is possible to tokenise the data and store it in a single ledger to unify the system of sharing and accessing IP systems.

## 6.2 Scalability value to the internet and industries

Enabling micro payment transactions

As the consensus mechanism does not rely heavily on rewarding the miners or staked pool holders, it will be possible to accommodate micro payment transactions cross chain and cross blockchain which is not possible today, even after ETH 2.0 [81] activates. The staked pool will have to be tipped which will increase the cost of the transactions. As shown in Figure 4, an average transaction fee for Ethereum as of August 2021 is at $3.31 and it went up to almost $75 in June 2021; this is not suitable for any micro transactions.

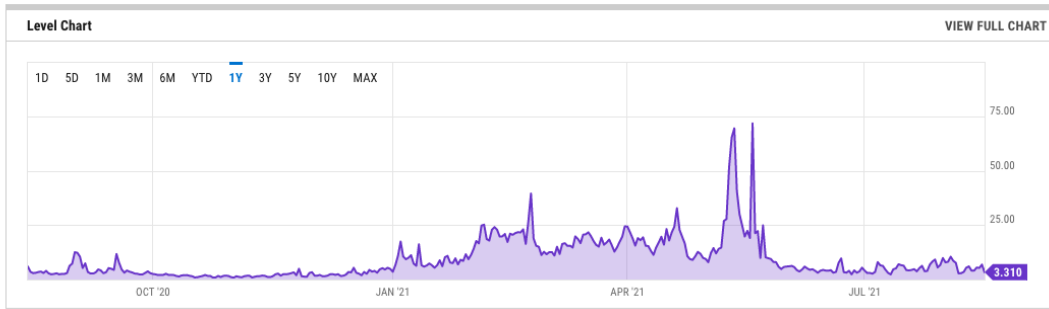

Figure 4. Average transaction fee for Ethereum between Sept 2020 and Aug 2021

Enabling scalable value exchange for cross blockchain

Whilst Polkadot adopts the delegated Proof-of-Stake consensus algorithm and makes use of bridges, parachain, and relay chain technologies, the cross blockchain ledger validation is governed strictly by the systems Polkadot provides. With Layer-One.S, the governing mechanisms will be indexed and abstracted by the layer two scaling solutions. The self-learning consensus mechanism will weigh the factors that develop into cross blockchain value or transaction exchange.

Transactional Use Case

Objective: To provide a single multi blockchain wallet mechanism for users to access the cryptocurrency landscape to buy, sell, stake, and earn interest in the native cryptocurrency whilst leveraging any decentralised staking pool crypto pairs.

Problem: To leverage DOT staking pools; ETH coins must be first exchanged with USDT and then with DOT coins. This increases the cost of the transaction around 15-20% cost to access DOT coins and the DOT staking pools charge around 20% on the interest that is earned as shown in Table 3 and Table 4 (Appendix I shows screenshots of the charging). To interoperate in between the tokens on the Ethereum network, the tokens must be recognised in the Ethereum ecosystem and only then the value can be exchanged.

Table 3. Comparison between conventional network with Layer-One network on overall cost

|  | Buy ETH | Contract Deployment | Exchange | Stake Interest |
|---|---|---|---|---|
| Conventional: ~25% value lost | Tx Fees | Smart Contract Fees | Exchange Fees | Interest Fees |
| Layer-One: ~1-2% cost | Internal Tx Fees | No Deployment Cost | Markup Commissions | Markup Commissions |

Table 4. Convention network fees and its percentages (Appendix I showing screenshots of Wyre fees, Network fees, and Contract Deployment fees)

| Conventional Network | | | | |
|---|---|---|---|---|
| Enter | Wyre Fees | Network Fees | Contract Deployment Fees | Total |
| $75.00 | $5.00 | $5.64 | $6.52 | $17.16 |
| % | 6.67% | 7.52% | 9% | 22.88% |

For conventional transaction on ETH, one needs to create a wallet and deploy a contract into Ethereum and then swap ETH with ERC20 Tokens which incur Gas Fees. Our solution will leverage the Layer-One Scalability and Interoperability solutions to manage a cross blockchain applicable wallet system that will integrate and exchange values in between different wallets without deploying the smart contract against every action that the user makes. It will manage the ledger using DAG and blockchain traversing through a custom data structure to permit a faster transaction throughput, lower transaction cost and instant finality.

Functionalities for the transactional use case are as follows and shown in Figure 5.

- Multi blockchain authenticated wallet to enable accessibility to DeFi liquidity pool and singular contract deployment.
- Avoiding multi-layer transactional costing.
- Facility to access liquidity pools across blockchain.
- Non-custodial facilitation of staking.
- Marginal mark-up and revenue fee model.

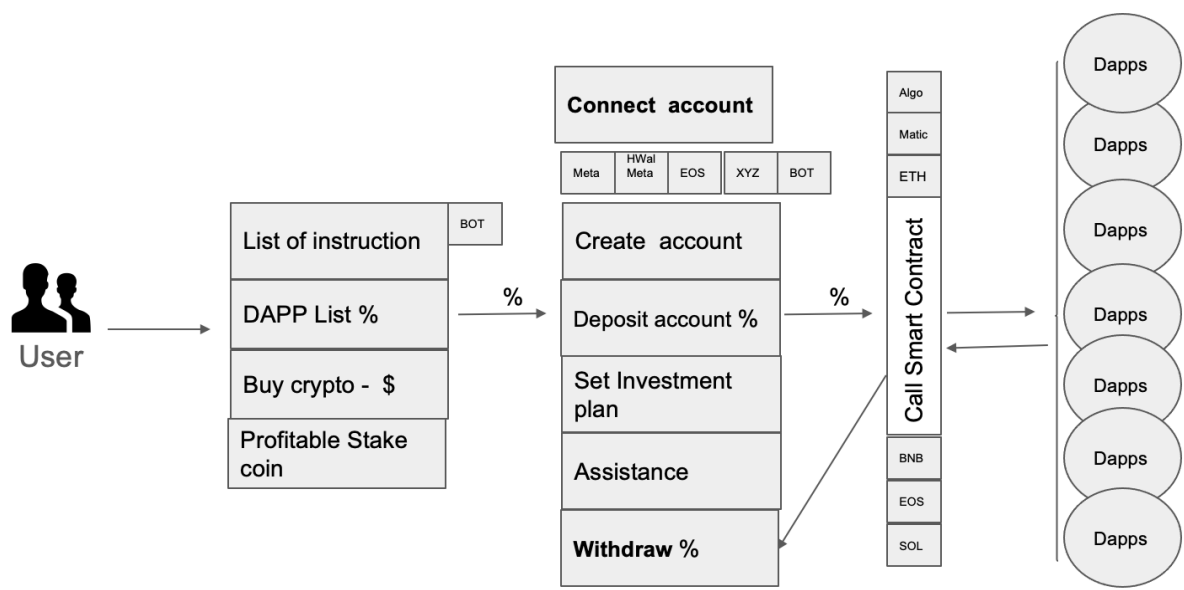

Figure 5. Schematic overview of transactional use case

Enabling micro trading for retail opportunity

Layer-One.X solution could decentralise trustless P2P network to connect buyers and sellers without a middleman, unnecessary fees or restrictions. Given the micro payment features enabled in the Layer-One.X solution, opportunities and efficiencies to the retail industry can be opened up.

For example, a renewable energy marketplace could utilise micro trading on a retail Layer-One.X solution to transact rooftop solar energy trading, down to 5-minute intervals, instead of 30-minute intervals through Power Ledger's blockchain platform [115] or 30/60-day billing cycle through Synergy (Government electricity supplier). The settlement period for the current wholesale Australian electricity market is every 30 minutes. However, starting in 2021, the settlement period will change to every five minutes [116].

Blockchain scalability solution has been empirically modelled to reflect blockchain trilemma for P2P energy trading [117] by processing all off chain transactions and recording them in a side chain. As a result, the block finality time can be set to bypass computation costs for every single transaction being recorded in the blockchain. Nevertheless, even though the results in greater cost saving, there could be a dispute occurring at the settlement phase due to the inconsistent state. With the Sync process and Nucleus script in the Layer-One.X solution, inconsistent states can be avoided at any phase.

[92]   J. Purdy, "Taxi Medallions: A Conceptual Framework For Work Tokens," Nov. 05, 2018. https://medium.com/messaricrypto/taxi-medallions-a-conceptual-framework-for-work-tokens-af69d581ac5d

[93]   K. Samani, "New Models for Utility Tokens," Feb. 13, 2018. https://multicoin.capital/2018/02/13/new-models-utility-tokens/

[94]   R. Holden and A. Malani, "The ICO paradox: Transactions costs, token velocity, and token value," National Bureau of Economic Research, 2019.

[95]   "The Augur White Paper: A Decentralized Oracle and Prediction Market Platform," Jan. 30, 2018. https://medium.com/@AugurProject/the-augur-white-paper-a-decentralized-oracle-and-prediction-market-platform-ed8907401c48

[96]   M. Luongo and C. Pon, "The keep network: A privacy layer for public blockchains," KEEP Network, Tech. Rep., 2018.[Online]. Available: https://keep. network …, 2017.

[97]   D. Petkanics, "Livepeer Whitepaper." Medium, Apr. 21, 2017. [Online]. Available: https://petkanics.medium.com/livepeer-whitepaper-9e8b88418e30

[98]   J. Koch and C. Reitwießner, "A predictable incentive mechanism for TrueBit," *ArXiv Prepr. ArXiv180611476*, 2018.

[99]   R. O'Reilly, "Introducing Gems: The Protocol for Decentralized Mechanical Turk." Nov. 03, 2017. [Online]. Available: https://medium.com/expand/introducing-gems-the-protocol-for-decentralized-mechanical-turk-8bd5ef29ca82

[100]  D. Vieira Fernandes, "Tokens,'Smart Contracts' and System Governance," *Smart Contracts Syst. Gov. Novemb. 23 2019*, 2019.

[101]  "The Maker Protocol: MakerDAO's Multi-Collateral Dai (MCD) System," Feb. 2020. https://makerdao.com/en/whitepaper/#addressable-market

[102]  A. Felix, "The Fundamentals of Discount Tokens," Feb. 20, 2018. https://blog.coinfund.io/the-fundamentals-of-discount-tokens-cc400c66198e

[103]  "Binance whitepaper." whitepaper.io. [Online]. Available: https://whitepaper.io/document/10/binance-whitepaper

[104]  "Sweetbridge whitepaper." [Online]. Available: https://sweetbridge.com/wp-content/uploads/2021/07/Sweetbridge-Whitepaper-20180529.pdf

[105]  N. Tsang, "A Look Back At The CoinFi Whitepaper And The Path Forward Through Crypto Winter." CoinFi.com, Mar. 05, 2019. [Online]. Available: https://docs.google.com/document/d/1p6xaFl4nPv1CuJv6F2fkZ6qBq2lBS6ePyFna8QZt1KM/edit

[106]  G. Adams, "Burn and Mint Equilibrium," Aug. 12, 2019. https://messari.io/article/burn-and-mint-equilibrium

[107]  "Factom, a data publishing layer atop major blockchains." https://messari.io/asset/factom

[108]  "Scriptarnica." https://medium.com/scriptarnica/latest

[109]  S. G. Gohwong, "The State of the Art of Cryptocurrencies," *Asian Adm. Manag. Rev.*, vol. 1, no. 2, 2018.

[110]  "Token, Golem." https://golem.network/

[111]  T. L. Penny, "Basic Attention Token & Brave: Digital Marketing on the Blockchain," 2018.

[112]  W. Warren and A. Bandeali, "0x: An open protocol for decentralized exchange on the Ethereum blockchain," *URl Httpsgithub Com0xProjectwhitepaper*, pp. 04–18, 2017.

[113]  K. S. Garewal, "The Helium Cryptocurrency Project," in *Practical Blockchains and Cryptocurrencies*, Springer, 2020, pp. 69–78.

[114]  T. McConaghy, "Towards a Practice of Token Engineering," Mar. 01, 2018. https://blog.oceanprotocol.com/towards-a-practice-of-token-engineering-b02feeeff7ca

[115]  "Fremantle Residents Participating In World-First Trial, Trading Solar Energy Peer-to-Peer and Setting Their Own Prices," Dec. 14, 2018. https://medium.com/power-ledger/fremantle-residents-participating-in-world-first-trial-trading-solar-energy-peer-to-peer-and-955b81d438c1
37

## Appendix I

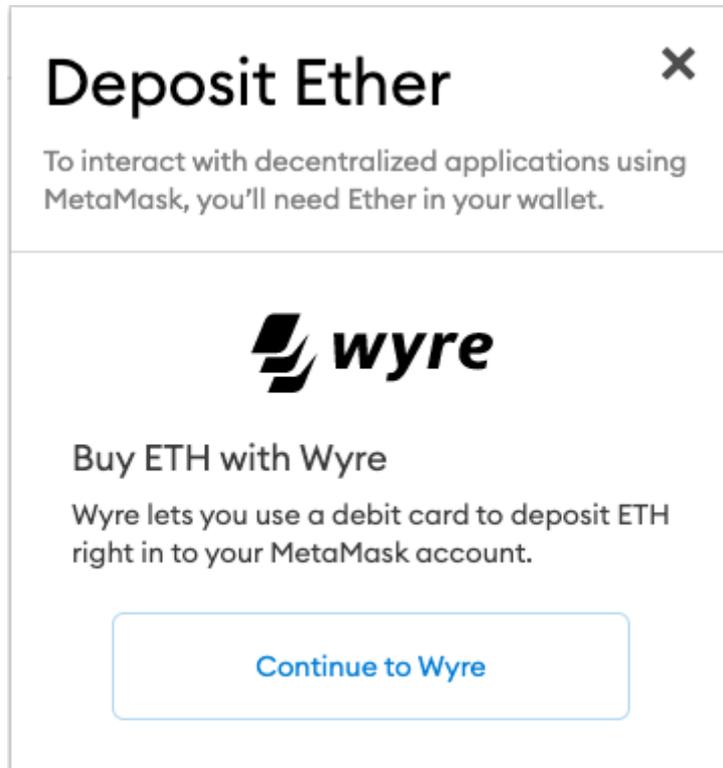

Figure A1. Wyre Fees

Figure A2. Network Fees

Figure A3. Contract Deployment Fees